\documentclass[reqno]{amsart}

\usepackage{booktabs}
\usepackage{IEEEtrantools}
\usepackage{amssymb,latexsym,amsfonts,amsmath}
\usepackage{graphicx}
\usepackage{mathrsfs}
\usepackage{dsfont}

\usepackage{tikz}
\usetikzlibrary{calc,shapes,arrows}

\usepackage{algorithm}
\usepackage{algorithmic}

\usepackage{xspace}

\newtheorem{theorem}{Theorem}[section]

\newtheorem{proposition}[theorem]{Proposition}

\newtheorem{definition}[theorem]{Definition}
\newtheorem{example}[theorem]{Example}

\newtheorem{remark}[theorem]{Remark}

\numberwithin{equation}{section}

\newcommand{\R}{{\mathbb{R}}}

\newcommand{\N}{{\mathbb{N}}}

\begin{document}

\begin{abstract}
	We propose a compositional approach for constructing abstractions of general Markov decision processes using approximate probabilistic relations. The abstraction framework is based on the notion of $\delta$-lifted relations, using which one can quantify the distance in probability between the interconnected gMDPs and that of their abstractions. This new approximate relation unifies compositionality results in the literature by incorporating the dependencies between state transitions explicitly and by allowing abstract models to have either finite or infinite state spaces. Accordingly, one can leverage the proposed results to perform analysis and synthesis over abstract models, and then carry the results over concrete ones. To this end, we first propose our compositionality results using the new approximate probabilistic relation which is based on lifting. We then focus on a class of stochastic nonlinear dynamical systems and construct their abstractions using both model order reduction and space discretization in a unified framework. We provide conditions for simultaneous existence of relations incorporating the structure of the network. Finally, we demonstrate the effectiveness of the proposed results by considering a network of four nonlinear dynamical subsystems (together 12 dimensions) and constructing finite abstractions from their reduced-order versions (together 4 dimensions) in a unified compositional framework. We benchmark our results against the compositional abstraction techniques that construct both infinite abstractions (reduced-order models) and finite MDPs in two consecutive steps. We show that our approach is much less conservative than the ones available in the literature.
\end{abstract}

\title[Compositional Abstraction-based Synthesis of MDPs via Approximate Probabilistic Relations]{Compositional Abstraction-based Synthesis of General MDPs via Approximate Probabilistic Relations}

\author{Abolfazl Lavaei$^1$}
\author{Sadegh Soudjani$^2$}
\author{Majid Zamani$^{3,4}$}
\address{$^1$Department of Electrical and Computer Engineering, Technical University of Munich, Germany.}
\email{lavaei@tum.de}
\address{$^2$School of Computing, Newcastle University, UK.}
\email{sadegh.soudjani@ncl.ac.uk}
\address{$^3$Department of Computer Science, University of Colorado Boulder, USA.}
\address{$^4$Department of Computer Science, Ludwig Maximilian University of Munich, Germany.}
\email{majid.zamani@colorado.edu}
\maketitle

\section{Introduction}
{\bf Motivations.} Control systems with stochastic uncertainty can be modeled as Markov decision processes (MDPs) over general state spaces. Synthesizing policies for satisfying complex temporal logic properties over MDPs evolving on uncountable state spaces is inherently a challenging task due to the computational complexity. Since closed-form characterization of such policies is not available in general, a suitable approach is to approximate these models by simpler ones possibly with finite or lower dimensional state spaces. A crucial step is to provide formal guarantees during this approximation phase, such that the analysis or synthesis on the simpler model can be refined back over the original one. In other words, one can first abstract the original model by a simpler one, and then carry the results from the simpler model to the concrete one using an interface map, by providing quantified errors on the approximation.

{\bf Related literature.} Similarity relations over finite-state stochastic systems have been studied, either via exact notions of probabilistic (bi)simulation relations~\cite{larsen1991bisimulation},~\cite{segala1995probabilistic} or approximate versions~\cite{desharnais2008approximate},~\cite{d2012robust}.
Similarity relations for models with general, uncountable state spaces have also been proposed in the literature. These relations either depend on stability requirements on model outputs via martingale theory or contractivity analysis~\cite{julius2009approximations},~\cite{zamani2014symbolic} or enforce structural abstractions of a model~\cite{desharnais2004metrics} by exploiting continuity conditions on
its probability laws~\cite{abate2013approximation},~\cite{abate2014probabilistic}.
These similarity relations are then used to relate the probabilistic behavior of a concrete model to that of its abstraction.
There have been also several results on the construction of (in)finite abstractions for stochastic systems. Construction of finite abstractions for formal verification and synthesis is presented in~\cite{APLS08}. Extension of such techniques to automata-based controller synthesis and infinite horizon properties, and improvement of the construction algorithms in terms of scalability are proposed in~\cite{Kamgarpour2013},~\cite{tkachev2011infinite}, and~\cite{SA13}, respectively.

In order to make the techniques applicable to networks of interacting systems, compositional abstraction and policy synthesis are studied in the literature. Compositional construction of finite abstractions using dynamic Bayesian networks is discussed in \cite{SAM15}. Compositional construction of infinite abstractions (reduced-order models) is proposed  in \cite{lavaei2017compositional,lavaei2018CDCJ} using small-gain type conditions and dissipativity-type properties of subsystems and their abstractions, respectively. 
Compositional construction of finite abstractions is studied in \cite{lavaei2018ADHS,lavaei2017HSCC}.
Compositional modeling and analysis for the safety verification of stochastic hybrid systems are investigated in~\cite{hahn2013compositional} in which random behaviour occurs only over the discrete components -- this limits their applicability to systems with continuous probabilistic evolutions. Compositional modeling of stochastic hybrid systems is discussed in~\cite{strubbe2006compositional} using communicating piecewise deterministic Markov processes that are connected through a composition operator. Recently, compositional synthesis of large-scale stochastic systems using a relaxed dissipativity approach is proposed in~\cite{lavaei2019NAHS}.

{\bf Our Contributions.} In our proposed framework, we consider the class of general Markov decisions processes (gMDPs), which evolves over continuous or uncountable state spaces, equipped with an output space and an output map. We encode interaction between gMDPs via \emph{internal} inputs, as opposed to \emph{external} inputs which are used for applying the synthesized policies enforcing some complex temporal logic properties.
We provide conditions under which the proposed similarity relations between individual gMDPs can be extended to relations between their respective interconnections.
These conditions enable compositional quantification of the distance in probability between the interconnected gMDPs and that of their abstractions.
The proposed notion has the advantage of encoding prior knowledge on dependencies between uncertainties of the two models. 
Our compositional scheme allows constructing both infinite and finite abstractions in a unified framework. We benchmark our results against the compositional abstraction techniques of~\cite{ lavaei2017HSCC,lavaei2018CDCJ} which are based on dissipativity-type reasoning and provide a compositional methodology for constructing both infinite abstractions (reduced-order models) and finite MDPs in two consecutive steps. We show that our approach is much less conservative than the ones proposed in~\cite{lavaei2017HSCC,lavaei2018CDCJ}. 

{\bf Recent Works.} Similarities between two gMDPs have been recently studied in~\cite{SIAM17} using a notion of {$\delta$-lifted relation}, but only for single gMDPs. The result is generalized in \cite{HSA18_ADHS} to a larger class of temporal properties and in \cite{SS18_robustDP} to synthesize policies for robust satisfaction of specifications.
One of the main contributions of this paper is to extend this notion such that it can be applied to networks of gMDPs. This extension is inspired by the notion of disturbance bisimulation relation proposed in \cite{mallik2016compositional}. In particular, we extend the notion of {$\delta$-lifted relation} for networks of gMDPs and show that under specific conditions systems can be composed while preserving the relation. This type of relations enables us to provide the probabilistic closeness guarantee between two interconnected gMDPs (cf. Theorem~\ref{closeness}). Furthermore, we provide an approach for the construction
of finite MDPs in a unified framework for a class of stochastic nonlinear dynamical systems, considered as gMDPs, whereas the construction scheme in~\cite{SIAM17} only handles the class of linear systems.

{\bf Organization.} The rest of the paper is organized as follows.
Section~\ref{sec:gMDPs} defines the class of general Markov decision processes with internal inputs and output maps.
Section~\ref{sec:lifting} presents first the notion of $\delta$-lifted relations over probability spaces and then the notion of lifting for gMDPs.
Section~\ref{sec:interconnected} provides compositional conditions for having the similarity relation between networks of gMDPs based on relations between their individual components.
Section~\ref{sec:abstractions} provides details of constructing finite abstractions for a network of stochastic nonlinear control systems, which is based on both model order reduction and space discretization in a unified framework, together with the similarity relations.
Finally, Section~\ref{sec:case_study} demonstrates the effectiveness of our approach on a numerical case study.

\section{General Markov Decision Processes}
\label{sec:gMDPs}

\subsection{Preliminaries and Notations}
In this paper, we work on Borel measurable spaces, i.e., $(X, \mathcal B(X))$, where $\mathcal B(X)$ is the Borel sigma algebra on $X$, and restrict ourselves to Polish spaces (i.e., separable and completely metrizable spaces). Given the measurable space  $(X, \mathcal B(X))$, a probability measure $\mathbb{P}$ defines the probability space $(X, \mathcal B(X), \mathbb{P})$. We denote the set of all probability
measures on $(X, \mathcal B(X))$ as $\mathcal{P}(X, \mathcal B(X))$. A map $f : S\rightarrow Y$ is measurable whenever it is Borel measurable.

The sets of nonnegative and positive
integers, and real numbers are denoted by $\mathbb N := \{0,1,2,\ldots\}$, $\mathbb N_{\ge 1} := \{1,2,3,\ldots\}$, and $\mathbb R$, respectively.
For column vectors $x_i \in \mathbb R^{n_i}$, $n_i\in \mathbb N_{\ge 1}$, and $i\in\{1,\ldots,N\}$,
we denote by $x = [x_1;\ldots;x_N]$ the corresponding column vector of dimension $\sum_i n_i$.
Given a vector $x\in\mathbb{R}^{n}$, $\Vert x\Vert$ denotes the Euclidean norm of $x$. 
The identity and zero matrices in $\mathbb R^{n\times{n}}$ are denoted by $\mathds{I}_n$ and $\mathbf{0}_{n\times n}$, respectively. The symbols $\mathbf{0}_n$ and $\mathds{1}_n$ denote the column vector in $\mathbb R^{n}$ with all elements equal to zero and one, respectively. 
A diagonal matrix in $\mathbb R^{N\times{N}}$ with diagonal entries $a_1,\ldots,a_N$ starting from the upper left corner is denoted by $\mathsf{diag}(a_1,\ldots,a_N)$. Given functions $f_i:X_i\rightarrow Y_i$,
for any $i\in\{1,\ldots,N\}$, their Cartesian product $\prod_{i=1}^{N}f_i:\prod_{i=1}^{N}X_i\rightarrow\prod_{i=1}^{N}Y_i$ is defined as $(\prod_{i=1}^{N}f_i)(x_1,\ldots,x_N)=[f_1(x_1);\ldots;f_N(x_N)]$. Given sets $X$ and $Y$, a relation $\mathscr{R}\subseteq X \times Y$ is a subset of the Cartesian product $X \times Y$ that relates $x \in X$ with $y \in Y$ if $(x, y) \in \mathscr{R}$, which is equivalently denoted by $x\mathscr{R}y$.

\subsection{General Markov Decision Processes}

In our framework, we consider the class of general Markov decision processes (gMDPs) that evolves over continuous or uncountable state spaces. This class of models generalizes the usual notion of MDP \cite{baier2008principles} by including internal inputs that are employed for composition ~\cite{lavaei2017HSCC}, and by adding an output space over which properties of interest are defined \cite{SIAM17}.

\begin{definition}
	\label{def:gMDP}
	A general Markov decision process (gMDP) is a tuple
	\begin{equation}
		\label{eq:dt-SCS}
		\Sigma =(X,W,U,\pi, T,Y,h)
	\end{equation}
	where 
	\begin{itemize}
		\item $X\subseteq \mathbb R^n$ is a Borel space as the state space of the system. We denote by $(X, \mathcal B (X))$ the measurable space with $\mathcal B (X)$  being  the Borel sigma-algebra on the state space;
		\item $W\subseteq \mathbb R^p$ is a Borel space as the \emph{internal} input space of the system; 
		\item $U\subseteq \mathbb R^m$ is a Borel space as the \emph{external} input space of the system;
		\item $\pi =  \mathcal B(X) \rightarrow [0,1]$ is the initial probability distribution;
		\item $T:\mathcal B(X)\times X\times W\times U\rightarrow[0,1]$ 
		is a conditional stochastic kernel that assigns to any $x \in X$, $w\in W$, and $\nu\in U$, a probability measure $T(\cdot | x,w,\nu)$
		on the measurable space $(X,\mathcal B(X))$. This stochastic kernel specifies probabilities over executions $\{x(k),k\in\mathbb N\}$ of the gMDP such that for any set $\mathcal{A} \in \mathcal B(X)$ and any $k\in\mathbb N$,
		\begin{align}\notag
			\mathbb P (x(k+1)\in \mathcal{A}\,\Big|\,& x(k),w(k),\nu(k))=  \int_\mathcal{A} T (dx(k+1)|x(k),w(k),\nu(k)).
		\end{align}
		\item $Y\subseteq \mathbb R^q$ is a Borel space as the output space of the system; 
		\item $h:X\rightarrow Y$ is a measurable function that maps a state $x\in X$ to its output $y = h(x)$.
	\end{itemize}
\end{definition}

\begin{remark}
	In this work, we are interested in networks of gMDPs that are obtained from composing gMDPs having both internal and external inputs and are synchronized through their internal inputs. The resulting interconnected gMDP will have only external input and will be denoted by the tuple $\Sigma =(X,U,\pi, T,Y,h)$  with stochastic kernel $T:\mathcal B(X)\times X\times U\rightarrow[0,1]$.
\end{remark}

Evolution of the state of a gMDP $\Sigma$, can be alternatively described by
\begin{equation}\label{Eq_1a}
	\Sigma:\left\{\hspace{-1.5mm}\begin{array}{l}x(k+1)=f(x(k),w(k),\nu(k),\varsigma(k)),\\
		y(k)=h(x(k)),\\
	\end{array}\right.
	\,\,\,\,\,\,k\in\mathbb N,\,\, x(0)\sim\pi,
\end{equation}
for input sequences $w(\cdot):\mathbb N\rightarrow W$ and $\nu(\cdot):\mathbb N\rightarrow U$,
where $\varsigma:=\{\varsigma(k):\Omega\rightarrow V_{\varsigma},\,\,k\in\mathbb N\}$ is a sequence of independent and identically distributed (i.i.d.) random variables on a set $V_\varsigma$ with sample space $\Omega$. Vector field $f$ together with the distribution of $\varsigma$ provide the stochastic kernel $T$.

The sets $\mathcal W$  and $\mathcal U$ are, respectively, associated to $W$ and $U$, collections of sequences $\{w(k):\Omega\rightarrow W,\,\,k\in\mathbb N\}$ and $\{\nu(k):\Omega\rightarrow U,\,\,k\in\mathbb N\}$, in which $w(k)$ and $\nu(k)$ are independent of $\varsigma(t)$ for any $k,t\in\mathbb N$ and $t\ge k$. For any initial state $a\in X$, $w(\cdot)\in\mathcal{W}$, $\nu(\cdot)\in\mathcal{U}$, the random sequence $y_{a w \nu}:\Omega \times \mathbb N \rightarrow Y$ satisfying~\eqref{Eq_1a} is called the \textit{output trajectory} of $\Sigma$ under initial state $a$, internal input $w$, and external input $\nu$. We eliminate subscript of $y_{a w \nu}$ wherever it is known from the context. If $ X, W, U$ are finite sets, system $\Sigma$ is called finite, and infinite otherwise.

Next section presents approximate probabilistic relations that can be used for relating two gMDPs while capturing probabilistic dependency between their executions. This new relation enables us to compose a set of concrete gMDPs and that of their abstractions while providing conditions for preserving the relation after composition.

\section{Approximate Probabilistic Relations based on Lifting}
\label{sec:lifting}

In this section, we first introduce the notion of  $\delta$-lifted relations over general state spaces. We then define ($\epsilon, \delta$)-approximate probabilistic relations based on lifting for gMDPs with internal inputs. Finally, we define ($\epsilon, \delta$)-approximate relations for interconnected gMDPs without internal input resulting from the interconnection of gMDPs having both internal and external inputs. First, we provide the notion of $\delta$-lifted relation borrowed from \cite{SIAM17}.

\begin{definition}
	\label{lifting}
	Let $X, \hat X$ be two sets with associated measurable spaces $(X, \mathcal B(X))$ and $(\hat X, \mathcal B(\hat X))$. Consider a relation
	$\mathscr{R}_x \in \mathcal B(X \times \hat X)$. We denote by $ \mathscr{\bar R}_{\delta}\subseteq \mathcal{P}(X, \mathcal B(X))\times \mathcal{P}(\hat X, \mathcal B(\hat X))$, the corresponding $\delta$-lifted relation if there exists a probability space $(X \times \hat X, \mathcal B(X \times \hat X), \mathscr{L})$ (equivalently, a lifting $\mathscr{L}$) such that $ (\Phi,\Theta)\in\mathscr{\bar R}_{\delta}$ if and only~if
	\begin{itemize}
		\item $\forall \mathcal{A} \in \mathcal B(X), ~\mathscr{L}(\mathcal{A} \times \hat X) = \Phi (\mathcal{A})$,
		\item $\forall \mathcal{\hat A} \in \mathcal B(\hat X), ~\mathscr{L}(X \times \mathcal{\hat A}) = \Theta (\mathcal{\hat A})$,
		\item for the probability space $(X \times \hat X, \mathcal B(X \times \hat X), \mathscr{L})$, it holds that $x\mathscr{R}_x \hat x$ with probability at least $1-\delta$,
		equivalently, $\mathscr{L}(\mathscr{R}_x)\geq 1-\delta$.
	\end{itemize}
\end{definition}

For a given relation $\mathscr{R}_x\subseteq X\times\hat X$, the above definition specifies required properties for lifting relation $\mathscr{R}_x$ to a relation $\mathscr{\bar R}_{\delta}$ that relates probability measures over $X$ and $\hat X$.

We are interested in using \emph {$\delta$-lifted} relation for specifying similarities between a gMDP and its abstraction. Therefore, internal inputs of the two gMDPs should be in a relation denoted by $\mathscr{R}_w$. Next definition gives conditions for having a stochastic simulation relation between two gMDPs.

\begin{definition}\label{Def: subsystems}
	Consider gMDPs $\Sigma =(X,W,U,\pi, T,Y,h)$ and $\widehat \Sigma =(\hat X,\hat W,\hat U,\hat \pi, \hat T,Y,\hat h)$ with the same output space. System $\widehat \Sigma$ is ($\epsilon, \delta $)-stochastically simulated by $\Sigma$, i.e. $ \widehat \Sigma\preceq_{\epsilon}^{\delta}\Sigma $, if there exist relations $\mathscr{R}_x\subseteq X \times \hat X$ and $\mathscr{R}_w\subseteq W \times \hat W$ for which there exists a Borel measurable stochastic kernel $\mathscr{L}_{T}(\cdot~|~ x, \hat x, w, \hat w, \hat \nu)$ on $X \times \hat X$ such that
	\begin{itemize}
		\item $\forall (x,\hat x) \in \mathscr{R}_x, ~\Vert h(x)- \hat h (\hat x) \Vert \leq \epsilon$, 
		\item $\forall (x,\hat x) \in \mathscr{R}_x$, $\forall \hat w \in \hat W$, $\forall \hat \nu \in \hat U$, there exists $\nu \in U$ such that $\forall  w \in  W$ with $(w,\hat w) \in \mathscr{R}_w $,
		$$T(\cdot~|~ x, w, \nu)~\mathscr{\bar R}_{\delta} ~ \hat T(\cdot~|~ \hat x, \hat w, \hat \nu)$$
		with lifting $\mathscr{L}_{T}(\cdot~|~ x, \hat x, w, \hat w, \hat \nu)$,
		\item $ \pi ~ \mathscr{\bar R}_{\delta}~ \hat \pi $.
	\end{itemize}
\end{definition}
Second condition of Definition~\ref{Def: subsystems} implies implicitly that there exists a function $\nu=\nu(x,\hat x,\hat w, \hat \nu)$ such that the state probability measures are in the lifted relation after one transition for any $(x,\hat x) \in \mathscr{R}_x$, $\hat w \in \hat W$, and $\hat \nu \in \hat U$. This function is called the \emph{interface function}, which can be employed for refining a synthesized policy $\hat\nu$ for $\widehat\Sigma$ to a policy $\nu$ for $\Sigma$. 

\begin{remark}\label{internal inputs}
	Definition~\ref{Def: subsystems} extends approximate probabilistic relation in \cite{SIAM17} by adding relation $\mathscr {R}_w$ to capture the effect of internal inputs. Interface function $\nu=\nu_{\hat \nu}(x,\hat x,\hat w, \hat \nu)$ is also allowed to depend on the internal input of the abstract gMDP~$\widehat\Sigma$.
\end{remark}

\begin{remark}
	\label{rem:joint}
	Note that Definition~\ref{Def: subsystems} generalizes the results of  \cite{lavaei2017compositional}, that assumes independent noises in two similar gMDPs,
	and of \cite{lavaei2017HSCC}, that assumes shared noises, by making no particular assumption but requiring this dependency to be reflected in lifting $\mathscr{L}_{T}$.
	We emphasize that this generalization is considered only for a concrete gMDP and its abstraction. We still retain the assumption of independent uncertainties between gMDPs in a network (cf.~Definition~\ref{def:comp} and Remark \ref{indep}).
\end{remark}

Definition~\ref{Def: subsystems} can be applied to gMDPs without internal inputs that may arise from composing gMDPs via their internal inputs. For such gMDPs, we eliminate $\mathscr{R}_w$ and interface function becomes independent of internal input, thus the definition reduces to that of \cite{SIAM17}, provided in the Appendix as Definition~\ref{Def: Interconnected}.

\begin{figure}[ht]
	\begin{center}
		\includegraphics[width=8.5cm]{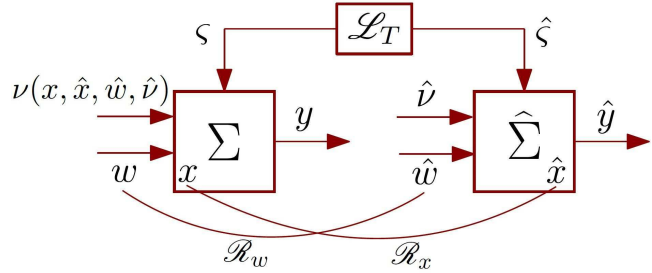}
		\caption{Notion of \emph {lifting} for specifying the similarity between gMDP and its abstraction. Relations $R_x$ and $R_w$ are the ones between states and internal inputs, respectively. $\mathscr{L}_{T}$ specifies the relation of two noises, and interface function $\nu_{\hat \nu}(x,\hat x,\hat w, \hat \nu)$ is used for the refinement policy.}
		\label{Fig1}
	\end{center}
\end{figure}

Figure~\ref{Fig1} illustrates ingredients of  Definition~\ref{Def: subsystems}. As seen, relation $R_w$ and stochastic kernel $\mathscr{L}_{T}$ capture the effect of internal inputs, and the relation of two noises, respectively. Moreover, interface function $\nu_{\hat \nu}(x,\hat x,\hat w, \hat \nu)$ is employed to refine a synthesized policy $\hat\nu$ for $\widehat\Sigma$ to a policy $\nu$ for $\Sigma$.

Definition~\ref{Def: subsystems} enables us to quantify the error in probability between a concrete system $\Sigma$ and its abstraction $\widehat\Sigma$. In any $(\epsilon, \delta)$-approximate probabilistic relation, $\delta $ is used to quantify the distance in probability between gMDPs and $\epsilon$ for the closeness of output trajectories as stated in the next theorem.
\begin{theorem}\label{closeness}
	If $ \widehat \Sigma\preceq_{\epsilon}^{\delta}\Sigma $ and $(w(k),\hat w(k)) \in \mathscr{R}_w$ for all $k\in\{0,1,\ldots,T_k\}$, then for all policies on $\widehat\Sigma$ there exists a policy for $\Sigma$ such that, for all measurable events $\mathsf{A} \subset Y^{T_k+1}$,
	\begin{align}\label{Closness}
		\mathbb{P}\{\{\hat y(k)\}_{0:T_k}\in \mathsf{A}^{-\epsilon}\}-\gamma &\leq \mathbb{P}\{\{y(k)\}_{0:T_k}\in \mathsf{A}\}\leq\mathbb{P}\{ \{\hat y(k)\}_{0:T_k}\in \mathsf{A}^{\epsilon}\}+\gamma,
	\end{align}
	with constant $1-\gamma := (1-\delta)^{T_k+1}$, and with the $\epsilon$-expansion and $\epsilon$-contraction of $\mathsf{A}$ defined as
	\begin{align*}
		\mathsf A^{\epsilon} &:=\{y(\cdot)\in Y^{T_k+1}\big|\exists \bar y(\cdot)\in \mathsf A\text{ with } max_{k\le T_k}\|\bar y(k)-y(k)\|\leq\epsilon\},\\
		\mathsf A^{-\epsilon} &:= \{y(\cdot)\in\mathsf A\,\,\big|\,\, \bar y(\cdot)\in \mathsf A\text{ for all } \bar y(\cdot) \text{ with } max_{k\le T_k}\|\bar y(k)-y(k)\|\leq\epsilon\}.
	\end{align*}
\end{theorem}

We have adapted this theorem from \cite{SIAM17} and added its proof in the Appendix for the sake of completeness. We employ this theorem to provide the probabilistic closeness guarantee between interconnected gMDPs and that of their compositional abstractions which are discussed in Section~\ref{sec:interconnected}.

In the next section, we define composition of gMDPs via their internal inputs and discuss how to relate them to a network of interconnected abstraction based on their individual relations.

\section{Interconnected gMDPs and Their Compositional Abstractions}
\label{sec:interconnected}

\subsection{Interconnected gMDPs}

Let $\Sigma$ be a network of $N\in\mathbb N_{\geq1}$ gMDPs
\begin{align}\label{Com: subsytems}
	\Sigma_i=(X_i,W_i,U_i,\pi_i, T_i,Y_i,h_i),\quad i\in \{1,\dots, N\}.
\end{align}
We partition internal input and output of $\Sigma_i$ as
\begin{align}\label{config1}
	w_i=[w_{i1};\ldots;w_{i(i-1)};w_{i(i+1)};\ldots;w_{iN}],~~~y_i=[y_{i1};\ldots;y_{iN}],
\end{align}
and also output space and function as
\begin{align}\label{config2}
	h_i(x_i)=[h_{i1}(x_i);\ldots;h_{iN}(x_i)],~~~Y_i=\prod_{j=1}^{N}Y_{ij}.
\end{align}
The outputs $y_{ii}$ are denoted as \emph{external} ones, whereas the outputs $y_{ij}$ with $i\neq j$ as \emph{internal} ones which are employed for interconnection by requiring $w_{ji} = y_{ij}$. This can be explicitly written using appropriate functions $g_i$ defined as
\begin{align}\label{eq:connection}
	w_i = g_i(x_1,\ldots,x_N):= \left[h_{1i}(x_1);\ldots;h_{(i-1)i}(x_{i-1});h_{(i+1)i}(x_{i+1});\ldots;h_{Ni}(x_N)\right]\!\!.
\end{align}
If there is no connection from $\Sigma_{i}$ to $\Sigma_j$, then the connecting output function is identically zero for all arguments, i.e., $h_{ij}\equiv 0$. Now, we define the \emph{interconnected gMDP} $\Sigma$ as follows.

\begin{figure}
	\begin{center}
		\includegraphics[width=7.7cm]{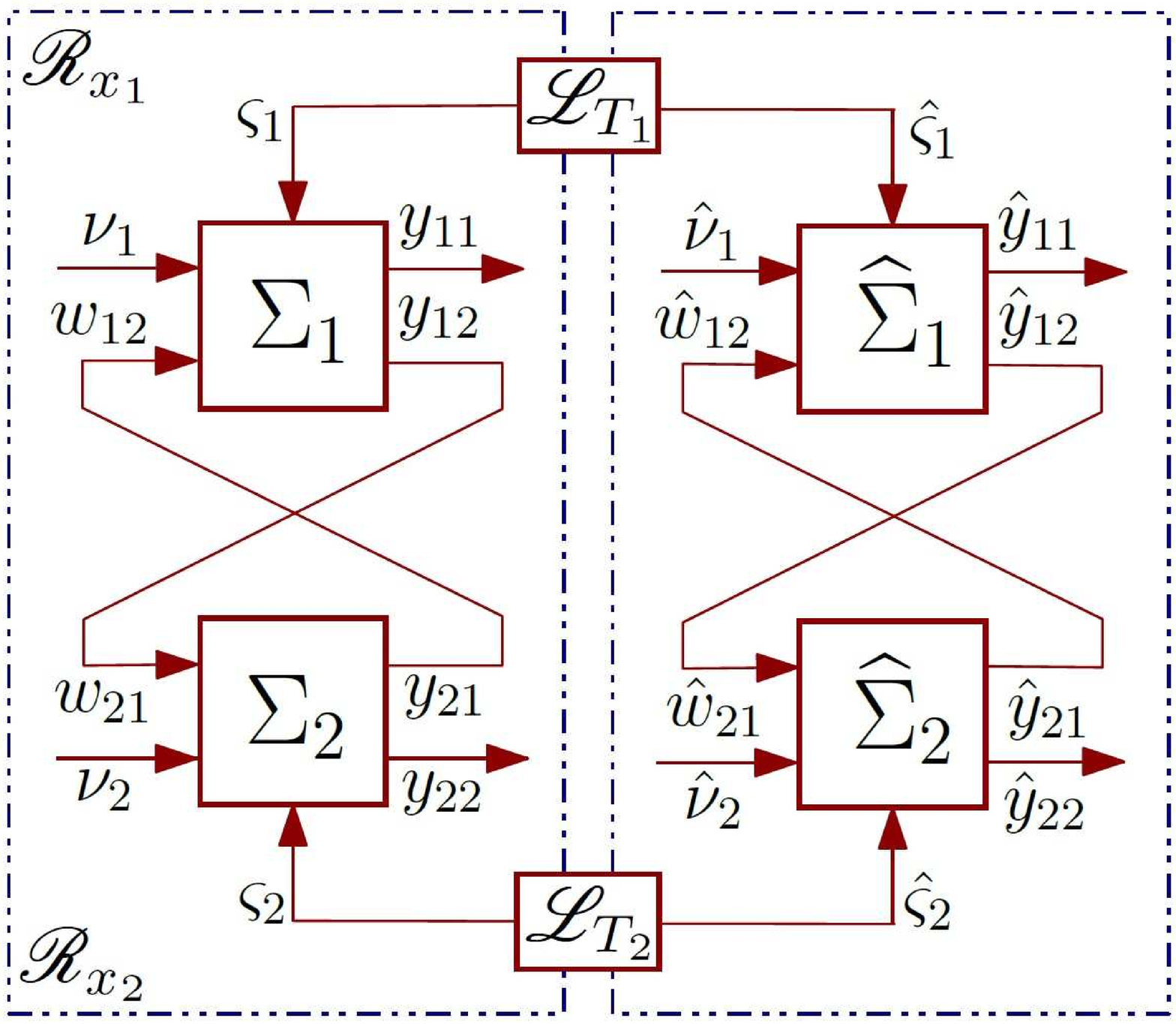}
		\caption{Interconnection of two gMDPs $\Sigma_1$ and $\Sigma_2$ and that of their abstractions.}
		\label{Fig2}
	\end{center}
\end{figure}

\begin{definition}
	\label{def:comp}
	Consider $N\in\N_{\geq1}$ gMDPs $\Sigma_i=(X_i,W_i,U_i,\pi_i, T_i,Y_i,h_i),
	i\in \{1,\dots, N\}$, with the input-output configuration as in \eqref{config1} and \eqref{config2}. The
	interconnection of  $\Sigma_i$, $i\in \{1,\dots, N\}$, is a gMDP $\Sigma=(X,U,\pi, T,Y,h)$, denoted by
	$\mathcal{I}(\Sigma_1,\ldots,\Sigma_N)$, such that $X:=\prod_{i=1}^{N}X_i$,  $U:=\prod_{i=1}^{N}U_i$, $Y:=\prod_{i=1}^{N}Y_{ii}$, and $h=\prod_{i=1}^{N}h_{ii}$, with the following constraints:
	\begin{align}
		\label{Inter constraint}
		\forall i,j\in \{1,\dots,N\},\,\,i\neq j: \quad w_{ji} = y_{ij}, \quad Y_{ij}\subseteq W_{ji}.
	\end{align}
	Moreover, one has conditional stochastic kernel $T:=\prod_{i=1}^{N}T_{i}$ and initial
	probability distribution $\pi:=\prod_{i=1}^{N}\pi_{i}$.
\end{definition}
An example of the interconnection of two gMDPs $\Sigma_1$ and  $\Sigma_2$ and that of their abstractions is illustrated in Figure~\ref{Fig2}.

\begin{remark}
	\label{indep}
	Definition~\ref{def:comp} assumes that uncertainties affecting individual gMDPs in a network $\mathcal{I}(\Sigma_1,\ldots,\Sigma_N)$ are independent and, thus, constructs $T$ and $\pi$ by taking products of $T_i$ and $\pi_i$, respectively. This definition can be generalized for dependent uncertainties by using their joint distribution in the construction of $T$ and $\pi$, in the same manner as we discussed in Remark~\ref{rem:joint} for expressing dependent uncertainties in concrete and abstract gMDPs.
\end{remark}

\subsection{Compositional Abstractions for Interconnected gMDPs}
We assume that we are given $N$ gMDPs as in Definition~\ref{def:gMDP} together with their corresponding abstractions $\widehat \Sigma_i=(\hat X_i,\hat W_i,\hat U_i,\hat \pi_i, \hat T_i,Y_i,\hat h_i)$ such that
$\widehat\Sigma_i\preceq_{\epsilon_i}^{\delta_i}\Sigma_i$ for some relation $\mathscr{R}_{x_i}$ and constants $\epsilon_{i},\delta_i$. 
Next theorem shows the main compositionality result of the paper.

\begin{theorem}\label{Thm1}
	Consider the interconnected gMDP
	$\Sigma=\mathcal{I}(\Sigma_1,\ldots,\Sigma_N)$ induced by $N\in\N_{\geq1}$ gMDPs~$\Sigma_i$. Suppose $\widehat \Sigma_i$ is ($\epsilon_i, \delta_i $)-stochastically simulated by $\Sigma_i$ with the corresponding relations $\mathscr{R}_{x_i}$ and $\mathscr{R}_{w_i}$ and lifting $\mathscr L_i$. If
	\begin{align}\label{compositionality condition}
		g_i(x)\mathscr R_{w_i} \hat g_i(\hat x), \quad\forall(x,\hat x)\in \mathscr{R}_{x_i},
	\end{align}  
	with interconnection constraint maps $g_i,\hat g_i$ defined as in \eqref{eq:connection}, then $\widehat \Sigma=\mathcal{I}(\widehat \Sigma_1,\ldots,\widehat\Sigma_N)$ is ($\epsilon, \delta $)-stochastically simulated by $\Sigma=\mathcal{I}(\Sigma_1,\ldots,\Sigma_N)$ with relation $\mathscr{R}_x$ defined as
	\begin{equation*}
		\begin{bmatrix}
			x_1\\
			\vdots\\
			x_N
		\end{bmatrix} 
		\mathscr{R}_x
		\begin{bmatrix}
			\hat x_1\\
			\vdots\\
			\hat x_N
		\end{bmatrix} 
		\Leftrightarrow
		\left\{\hspace{-1.5mm}\begin{array}{l}x_1	\mathscr{R}_{x_1}\hat x_1,\\
			~~~~~\vdots\\
			x_N	\mathscr{R}_{x_N}\hat x_N,
		\end{array}\right.
	\end{equation*}
	and constants
	$\epsilon = \sum_{i=1}^N \epsilon_{i}$, and $\delta = 1-\prod_{i=1}^{N}(1-\delta_i)$.
	Lifting $\mathscr L$ and interface $\nu$ are obtained by taking products $\mathscr L = \prod_{i=1}^N \mathscr L_i$ and $\nu = \prod_{i=1}^N \nu_i$, and then substituting interconnection constraints~\eqref{Inter constraint}.
\end{theorem}
The proof of Theorem~\ref{Thm1} is provided in the Appendix.

\begin{remark}
	Note that Theorem~\ref{Thm1} requires $g_i(x)\mathscr R_{w_i} \hat g_i(\hat x)$ for any $(x,\hat x)\in \mathscr{R}_{x}$. This condition puts restriction on the structure of the network and how the dynamics of gMDPs are coupled in the network (cf. Remark~\ref{internal inputs}). It is similar to the condition imposed in disturbance bisimulation relation defined in \cite{mallik2016compositional}.
	
\end{remark}
We provide the following example to illustrate our compositionality results.
\begin{example} 
	Assume that we are given two linear dynamical systems as		
	\begin{align}\label{exm: com}
		\Sigma_i:\left\{\hspace{-1.5mm}\begin{array}{l}x_i(k+1)=A_ix_i(k)+D_iw_i(k)+B_i\nu_i(k)+R_i\varsigma_i(k),\\
			y_i(k)=x_i(k),\quad
			i\in\{1,2\},
		\end{array}\right.
	\end{align}
	where the additive noise $\varsigma_i(\cdot)$ is a sequence of independent random vectors with multivariate standard normal distributions for $i \in\{1,2\}$, and $R_i,i\in\{1,2\},$ are invertible. Let $\widehat\Sigma_i$ be the abstraction of gMDP~\eqref{exm: com} as
	\begin{align}\notag
		\widehat \Sigma_i:\left\{\hspace{-1.5mm}\begin{array}{l}\hat x_i(k+1)=\hat A_i \hat x_i(k)+\hat D_i \hat w_i(k)+\hat B_i \hat\nu_i(k)+\hat R_i \hat \varsigma_i(k),\\
			\hat y_i(k)=\hat x_i(k).\end{array}\right.
	\end{align}
	Transition kernels of $\Sigma_i$ and $\widehat \Sigma_i$ can be written as 
	\begin{align}
		\notag
		&T_i(\cdot| x_i, w_i, \nu_i) = \mathcal{N}(\cdot|A_ix_i+D_iw_i+B_i\nu_i, R_iR_i^T),\\\notag
		&\hat T_i(\cdot| \hat x_i, \hat w_i, \hat \nu_i) = \mathcal{N}(\cdot|\hat A_i \hat x_i+\hat D_i \hat w_i+\hat B_i \hat \nu_i, \hat R_i \hat R_i^T), ~\forall i \in \{1,2\},
	\end{align}
	where $\mathcal{N}(\cdot\,|\, \mathsf m,\mathsf D)$ indicates normal distribution with mean $\mathsf m$ and covariance matrix $\mathsf D$.
	
	\noindent\textbf{Independent uncertainties.} If $\varsigma_i(\cdot)$ and $\hat \varsigma_i(\cdot)$ in the concrete and abstract systems are independent, a candidate for lifted measure is
	\begin{align}\notag
		\mathscr{L}_{T_i}(\cdot| x_i, \hat x_i, w_i, \hat w_i, \hat \nu_i)&= \mathcal{N}(\cdot|A_ix_i+D_iw_i+B_i\nu_i, R_iR_i^T)\times\mathcal{N}(\cdot|\hat A_i \hat x_i+\hat D_i \hat w_i+\hat B_i \hat \nu_i, \hat R_i \hat R_i^T).
	\end{align}
	Now we connect two subsystems with each other based on the interconnection constraint \eqref{Inter constraint} which are $w_i = x_{3-i}$ and $\hat w_i = \hat x_{3-i}$ for $i\in\{1,2\}$. 
	For any $x=[{x_1;x_2}]\in X, \hat x=[{\hat x_1;\hat x_2}]\in \hat X, \nu=[{\nu_1;\nu_2}]\in U, \hat \nu=[{\hat \nu_1;\hat \nu_2}]\in \hat U$,
	the compositional transition kernels for the interconnected gMDPs are
	\begin{align}\notag
		T(\cdot~|~ x, \nu) = \mathcal{N}(\cdot~|~Ax+B\nu, RR^T),~\hat T(\cdot~|~ \hat x, \hat \nu) = \mathcal{N}(\cdot~|~\hat A \hat x+\hat B \hat \nu, \hat R \hat R^T),
	\end{align}
	where $\nu:=\nu(x,\hat x, \hat \nu)$ and
	\begin{align}\notag
		&A= \begin{bmatrix}
			A_1 & D_1\\
			D_2 & A_2
		\end{bmatrix}\!\!,
		\quad B=\mathsf{diag}(B_1,B_2),
		\quad R=\mathsf{diag}(R_1,R_2),\\
		&\hat A= \begin{bmatrix}
			\hat A_1 & \hat D_1\\
			\hat D_2 & \hat A_2
		\end{bmatrix}\!\!,
		\quad\hat B=\mathsf{diag}(\hat B_1,\hat B_2),
		\quad\hat R=\mathsf{diag}(\hat R_1,\hat R_2).
		\label{eq:matrices}
	\end{align}
	Then the candidate lifted measure for the interconnected gMDPs is
	\begin{align}\notag
		&\mathscr{L}_{T}(\cdot| x, \hat x,\hat \nu) =\mathcal{N}(\cdot|Ax+B\nu, RR^T)\mathcal{N}(\cdot|\hat A \hat x+\hat B \hat \nu, \hat R \hat R^T).
	\end{align}
	
	Note that after connecting the subsystems with each other using the proposed interconnection constraint in~\eqref{Inter constraint}, the internal inputs will disappear. 
	
	\smallskip
	\noindent\textbf{Dependent uncertainties.}
	Suppose $\Sigma_i$ and $\widehat\Sigma_i$ share the same noise $\varsigma_i(\cdot) = \hat\varsigma_i(\cdot)$. 
	In this case, the candidate lifted measure for $i \in \{1,2\}$ is obtained by
	\begin{align*}
		\mathscr{L}_{T_i}(dx_i'\times  d\hat x_i' | x_i, \hat x_i, w_i, \hat w_i, \hat \nu_i)&=\mathcal{N}(d x_i' | A_ix_i+D_iw_i+B_i\nu_i,R_iR_i^T)\times\delta_d(d\hat x_i' |\hat A_i \hat x_i\\
		&+\hat D_i \hat w_i+\hat B_i \hat \nu_i+\hat R_i R_i^{-1}(x_i'-A_ix_i-D_iw_i-B_i\nu_i)),
	\end{align*}
	where $\delta_d(\cdot|\mathsf a)$ indicates Dirac delta distribution centered at $\mathsf a$.
	Now we connect two subsystems with each other.  For any $x=[{x_1;x_2}]\in X, \hat x=[{\hat x_1;\hat x_2}]\in \hat X, \nu=[{\nu_1;\nu_2}]\in U, \hat \nu=[{\hat \nu_1;\hat \nu_2}]\in \hat U$, 
	the candidate lifted measure for the interconnected gMDPs is
	\begin{align*}
		\mathscr{L}_{T}(dx'\times d\hat x'| x, \hat x,\hat \nu) =\mathcal{N}&(dx' | Ax+B\nu, RR^T)\times\delta_d(d\hat x'|A \hat x
		+ B \hat \nu-\bar A x+\tilde A x'-\bar B \nu),
	\end{align*}
	where $A,B,R,\hat A,\hat B$ are defined as in \eqref{eq:matrices}, and
	\begin{align}\notag
		\bar A= \begin{bmatrix}
			\hat R_1 R_1^{-1} A_1 & \hat R_1 R_1^{-1} D_1\\
			\hat R_2 R_2^{-1} D_2 & \hat R_2 R_2^{-1} A_2\\
		\end{bmatrix}\!\!,
		\quad\tilde A= \begin{bmatrix}
			\hat R_1 R_1^{-1} & 0\\
			0 & \hat R_2 R_2^{-1}\\
		\end{bmatrix}\!\!,\quad\bar B= \begin{bmatrix}
			\hat R_1 R_1^{-1} B_1 & 0\\
			0 & \hat R_2 R_2^{-1} B_2\\
		\end{bmatrix}\!\!.
	\end{align}
\end{example}

In the next section, we focus on a particular class of stochastic nonlinear systems, and construct its infinite and finite abstractions in a unified framework. We provide explicit inequalities for establishing Theorem~\ref{Thm1}, which gives a probabilistic relation after composition and enables us to get guarantees of Theorem~\ref{closeness} on the closeness of the composed system and that of its abstraction.

\section{Construction of Abstractions for Nonlinear Systems}
\label{sec:abstractions}

Here, we focus on a specific class of stochastic nonlinear control systems $\Sigma$ as

\begin{align}\label{Eq_58a}
	\Sigma:\left\{\hspace{-1.5mm}\begin{array}{l}x(k+1)=Ax(k)+E\varphi(Fx(k))+Dw(k)+B\nu(k)+R\varsigma(k),\\
		y(k)=Cx(k),\end{array}\right.
\end{align}
where $\varsigma(\cdot)\sim\mathcal N(0, \mathds{I}_n)$, and $\varphi:\R\rightarrow\R$ satisfies 
\begin{equation}\label{Eq_6a}
	a\leq\frac{\varphi(c)-\varphi(d)}{c-d}\leq b,~~~\forall c,d\in\R,c\neq d,
\end{equation}
for some $a\in\R$ and $b\in\R_{>0}\cup\{\infty\}$, $a\leq b$. 

We use the tuple
\begin{align}\notag
	\Sigma=(A,B,C,D,E,F,R,\varphi),
\end{align}
to refer to the class of nonlinear systems of the form~\eqref{Eq_58a}.
\begin{remark}
	If $E$ is a zero matrix or $\varphi$ in~\eqref{Eq_58a} is linear including the zero function (i.e. $\varphi\equiv0$), one can remove or push the term $E\varphi(Fx)$ to $Ax$, and consequently the nonlinear tuple reduces to the linear one $\Sigma=(A,B,C,D,R)$. Then, every time we mention the tuple $\Sigma=(A,B,C,D,E,F,R,\varphi)$, it implicitly implies that $\varphi$ is nonlinear and $E$ is nonzero. 
\end{remark}

\begin{remark}
	Without loss of generality~\cite{arcak2001observer}, we can assume $a=0$ in~\eqref{Eq_6a} for the class of nonlinear systems in~\eqref{Eq_58a}. If $a\neq0$, one can define a new function $\tilde\varphi(s):=\varphi(s)-as$ satisfying~\eqref{Eq_6a} with $\tilde a=0$ and $\tilde b=b-a$, and rewrite~\eqref{Eq_58a} as
	\begin{align}\notag
		\Sigma:\left\{\hspace{-1.5mm}\begin{array}{l}x(k+1)=\tilde Ax(k)+E\tilde \varphi(Fx(k))+Dw(k)+B\nu(k)+R\varsigma(k),\\
			y(k)=Cx(k)\end{array}\right.
	\end{align}
	where $\tilde A=A+aEF$.
\end{remark}

\begin{remark}\label{Re_1a}	
	We restrict ourselves here to systems with a single nonlinearity as in~\eqref{Eq_58a} for the sake of simple presentation. However, it would be straightforward to get analogous results for systems with multiple nonlinearities as
	\begin{align}\notag
		\Sigma:\left\{\hspace{-1.5mm}\begin{array}{l}x(k+1)=Ax(k)+\sum_{i=1}^{\bar M}E_i \varphi_i(F_ix(k))+Dw(k)+B\nu(k)+R\varsigma(k),\\
			y(k)=Cx(k),\end{array}\right.
	\end{align}
	where $\varphi_i:\R\rightarrow\R$ satisfies~\eqref{Eq_6a} for some $a_i\in\R$ and $b_i\in\R_{>0}\cup\{\infty\}$, for any $i\in\{1,\ldots,\bar M\}$.
\end{remark}

Existing compositional abstraction results for this class of models are based on either model order reduction~\cite{lavaei2017compositional},~\cite{lavaei2018CDCJ} or finite MDPs~\cite{lavaei2017HSCC},~\cite{lavaei2018ADHS}. Our proposed results here combine these two approaches in one unified framework. In other words, our abstract model is obtained by discretizing the state space of a reduced-order version of the concrete model.

\subsection{Construction of Finite Abstractions}
\label{subsec:finite}
Consider a nonlinear system $\Sigma=(A,B,C,D,E,F,R,\varphi)$ and its reduced-order version $\widehat\Sigma_{\textsf r}=(\hat A_{\textsf r},\hat B_{\textsf r},\hat C_{\textsf r},\hat D_{\textsf r},\hat E_{\textsf r},\hat F_{\textsf r},\hat R_{\textsf r}, \varphi)$. Note that index ${\textsf r}$ in the whole paper signifies the reduced-order version of the original model. We discuss the construction of $\widehat\Sigma_{\textsf r}$ from $\Sigma$ in Theorem~\ref{Thm2} of the next subsection. Construction of a finite gMDP from $\widehat\Sigma_{\textsf r}$ follows the approach of \cite{SSoudjani,SA13}. Denote the state and input spaces of $\widehat\Sigma_{\textsf r}$ respectively by $\hat X_{\textsf r}, \hat W_{\textsf r}, \hat U_{\textsf r}$.
We construct a finite gMDP by selecting partitions $\hat X_{\textsf r} = \cup_i \mathsf X_i$, $\hat W_{\textsf r} = \cup_i \mathsf W_i$, and $\hat U_{\textsf r} = \cup_i \mathsf U_i$, and choosing representative points $\bar x_i\in \mathsf X_i$, $\bar w_i\in \mathsf W_i$, and $\bar \nu_i\in \mathsf U_i$, as abstract states and inputs. 
The finite abstraction of $\Sigma$ is a gMDP $\widehat\Sigma= (\hat X,\hat W,\hat U,\hat \pi, \hat T,Y,\hat h)$,	
where
\begin{align}\notag
	&\hat X = \{\bar x_i,i=1,\ldots,n_x\},~ \hat U = \{\bar u_i,i=1,\ldots,n_u\},~\hat W = \{\bar w_i,i=1,\ldots,n_w\}.
\end{align}
Transition probability matrix $\hat T$ is constructed according to the dynamics $\hat x(k+1) = \hat f(\hat{x}(k),\hat{w}(k),\hat{\nu}(k),\varsigma(k))$ with
\begin{equation}
	\label{eq:abs_dyn}
	\hat f(\hat{x},\hat{\nu},\hat{w},\varsigma) := \Pi_x(\hat A_{\textsf r}\hat x+\hat E_{\textsf r}\varphi(\hat F_{\textsf r}\hat x)+\hat D_{\textsf r}\hat w+\hat B_{\textsf r}\hat \nu+\hat R_{\textsf r}\varsigma),	
\end{equation}
where $\Pi_x:\hat X_{\textsf r}\rightarrow \hat X$ is the map that assigns to any $\hat x_{\textsf r}\in \hat X_{\textsf r}$, the representative point $\hat x\in\hat X$ of the corresponding partition set containing $\hat x_{\textsf r}$. The output map $\hat h(\hat x) = \hat C \hat x$. The initial state of $\widehat\Sigma$ is also selected according to $\hat x_0 := \Pi_x(\hat x_{\textsf r}(0))$ with $\hat x_{\textsf r}(0)$ being the initial state of $\widehat\Sigma_{\textsf r}$.

\begin{remark}
	Abstraction map $\Pi_x$ satisfies the inequality
	$\Vert \Pi_x(\hat x_{\textsf r})-\hat x_{\textsf r}\Vert \leq \beta$ for all $\hat x_{\textsf r}\,\in \hat X_{\textsf r},$
	where $\beta$ is the state discretization parameter defined as $\beta:=\sup\{\|\hat x_{\textsf r}-\hat x_{\textsf r}'\|,\,\, \hat x_{\textsf r},\hat x_{\textsf r}'\in \mathsf X_i,\,i=1,2,\ldots,n_x\}$.
\end{remark}

\subsection{Establishing Probabilistic Relations}
In this subsection, we provide conditions under which $\widehat \Sigma$ is ($\epsilon, \delta $)-stochastically simulated by $\Sigma$, i.e. $ \widehat \Sigma\preceq_{\epsilon}^{\delta}\Sigma $, with relations $\mathscr{R}_x$ and $\mathscr{R}_w$. Here we candidate relations
\begin{IEEEeqnarray}{rCl}\IEEEyesnumber
	\IEEEyessubnumber\label{Eq10a} \mathscr{R}_x&=&\Big\{(x,\hat x)|(x-P\hat x)^TM(x-P\hat x)\leq \epsilon^2\Big\},\\
	\IEEEyessubnumber\label{Eq100a} \mathscr{R}_w&=&\Big\{(w,\hat w)|(w-P_w\hat w)^TM_w(w-P_w\hat w)\leq \epsilon_w^2\Big\},
\end{IEEEeqnarray}
where $P\in \mathbb R^{n\times\hat n}$ and $P_w\in \mathbb R^{m\times\hat m}$ are matrices of appropriate dimensions (potentially with the lowest $\hat n$ and $\hat m$), and $M, M_w$ are positive-definite matrices.

Next theorem gives conditions for having $ \widehat \Sigma\preceq_{\epsilon}^{\delta}\Sigma $ with relations \eqref{Eq10a} and \eqref{Eq100a}.

\begin{theorem}\label{Thm2}
	Let $\Sigma=(A,B,C,D,E,F,R,\varphi)$ and $\widehat\Sigma_{\textsf r}=(\hat A_{\textsf r},\hat B_{\textsf r},\hat C_{\textsf r},\hat D_{\textsf r},\hat E_{\textsf r},\hat F_{\textsf r},\hat R_{\textsf r}, \varphi)$ be two nonlinear systems with the same additive noise. Suppose $\widehat\Sigma$ is a finite gMDP constructed from $\widehat\Sigma_{\textsf r}$ according to subsection~\ref{subsec:finite}. Then $\widehat \Sigma$ is ($\epsilon, \delta $)-stochastically simulated by $\Sigma$  with relations \eqref{Eq10a}-\eqref{Eq100a} if 
	there exist matrices $K$, $Q$, $S$, $L_1$, $L_2$  and $\tilde R$ such that
	\begin{IEEEeqnarray}{rCl}\IEEEyesnumber
		\IEEEyessubnumber\label{Eq11a} M&\succeq&C^TC ,\\
		\IEEEyessubnumber\label{Eq12a} \hat C_{\textsf r}&=&CP,\\
		\IEEEyessubnumber\label{Eq13a}  \hat F_{\textsf r}&=&FP,\\
		\IEEEyessubnumber\label{Eq14a}  E&=&P\hat E_{\textsf r}-B(L_1-L_2),\\
		\IEEEyessubnumber\label{Eq15a} AP&=&P\hat A_{\textsf r}-BQ,\\
		\IEEEyessubnumber\label{Eq16a} DP_w&=&P\hat D_{\textsf r}-BS,\\
		\IEEEyessubnumber \label{Eq17a} \mathbb{P}\{(H+PG)^TM(H&+&PG)\le \epsilon^2\} \succeq 1-\delta,
	\end{IEEEeqnarray}
	where
	\begin{align}\notag
		H &= ((A+BK)+\bar\delta(BL_1+E)F)(x-P\hat x)+D(w-P_w\hat w)+(B\tilde R-P\hat B_{\textsf r})\hat \nu+(R-P\hat R_{\textsf r})\varsigma,\\\notag
		G &= \hat A_{\textsf r} \hat x+\hat E_{\textsf r}\varphi(\hat F_{\textsf r}\hat x)+\hat D_{\textsf r} \hat w+\hat B_{\textsf r} \hat \nu+\hat R_{\textsf r} \varsigma- \Pi_x(\hat A_{\textsf r} \hat x+\hat E_{\textsf r}\varphi(\hat F_{\textsf r}\hat x)+\hat D_{\textsf r} \hat w+\hat B_{\textsf r} \hat \nu+\hat R_{\textsf r} \varsigma). 
	\end{align}
\end{theorem}

The proof of Theorem~\ref{Thm2} is provided in the Appendix.
\begin{remark}
	\label{rem:chi}
	Note that condition \eqref{Eq17a} is a chance constraint. We satisfy this condition by selecting constant $c_\varsigma$ such that $\mathbb{P}\{\varsigma^T  \varsigma\le  c_\varsigma^2\} \geq 1-\delta$, and requiring $(H+PG)^TM(H+PG)\le \epsilon^2$ for any $\varsigma$ with $\varsigma^T  \varsigma\le  c_\varsigma^2$.
	Since $\varsigma \sim (0, \mathds{I}_n)$, $\varsigma^T \varsigma $ has chi-square distribution with $2$ degrees of freedom. Thus, $c_\varsigma = \mathcal{X}_2^{-1}(1-\delta)$ with $\mathcal{X}_2^{-1}$ being chi-square inverse cumulative distribution function with $2$ degrees of freedom.
\end{remark}

\section{Case Study}
\label{sec:case_study}
\begin{figure}
	\centering
	\begin{tikzpicture}[auto, node distance=2cm, >=latex]
	\tikzset{block/.style    = {thick,draw, rectangle, minimum height = 3em, minimum width = 3em}}
	
	\node[block] (sys3) at (3,0) {$\Sigma_3$};
	\node[block] (sys4) at (3,-1.5) {$\Sigma_4$};
	\node[block] (sys1) at (0,0) {$\Sigma_1$};
	\node[block] (sys2) at (0,-1.5) {$\Sigma_2$};
	
	\draw[->] ($(sys3.east)+(0,0.25)$) -- node[near end,below] {$y_{33}$} ++(1.5,0);
	
	\draw[->] ($(sys4.east)+(0,-.25)$) -- node[near end, above] {$y_{44}$} ++(1.5,0);
	
	\draw[<-] ($(sys1.west)+(0,-.25)$) -- node[below, near end] {$\nu_{1}$} ++(-0.55,0);
	
	\draw[<-] ($(sys2.west)+(0,0.25)$) -- node[above, near end] {$\nu_{2}$} ++(-0.55,0);
	
	\draw[<-] ($(sys3.west)+(0,-.25)$) -- node[below, near end] {$\nu_{3}$} ++(-0.55,0);
	
	\draw[<-] ($(sys4.west)+(0,0.25)$) -- node[above, near end] {$\nu_{4}$} ++(-0.55,0);
	
	\draw[->] ($(sys3.east)+(.5,.25)$) |- node[near start, right] {$y_{31}$}  ($(sys1.west)+(-.5,.75)$) |- ($(sys1.west)+(0,.25)$) ;
	\draw[fill] ($(sys3.east)+(.5,.25)$) circle (1pt);
	
	\draw[->] ($(sys4.east)+(.5,-.25)$) |-node[near start, right] {$y_{42}$} ($(sys2.west)+(-.5,-.75)$) |- ($(sys2.west)+(0,-.25)$);
	\draw[fill] ($(sys4.east)+(.5,-.25)$) circle (1pt);
	
	\draw[->] ($(sys1.east)$)              -- node[above] {$y_{14}$} 
	($(sys1.east)+(0.5,0.00)$)  --
	($(sys4.west)+(-.5,0)$) -- (sys4.west);
	
	\draw[->] ($(sys2.east)$)              -- node[below] {$y_{23}$} 
	($(sys2.east)+(0.5,0.00)$)  --
	($(sys3.west)+(-.5,0)$) -- (sys3.west);
	
	\end{tikzpicture}
	\caption{The interconnected gMDP $\Sigma=\mathcal{I}(\Sigma_1,\Sigma_2,\Sigma_3,\Sigma_4)$.}\label{Fig3}
\end{figure}
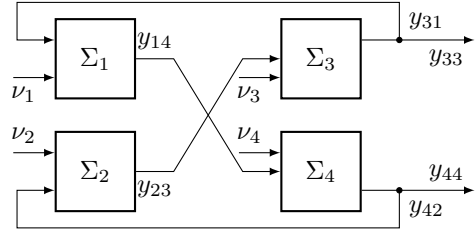

In this section, we demonstrate the effectiveness of the proposed results on a network of four stochastic nonlinear systems (totally 12 dimensions), i.e. $\Sigma=\mathcal{I}(\Sigma_1,\Sigma_2,\Sigma_3,\Sigma_4)$. We want to construct finite gMDPs from their reduced-order versions (together 4 dimensions). The interconnected gMDP $\Sigma$ is illustrated in Figure~\ref{Fig3} such that the output of $\Sigma_1$ (resp. $\Sigma_2$) is connected to the internal input of $\Sigma_{4}$ (resp. $\Sigma_3$), and the output of $\Sigma_{3}$ (resp. $\Sigma_4$) connects to the internal input of $\Sigma_{1}$ (resp. $\Sigma_2$).

The matrices of the system are given by
\begin{align}\notag
	&A_i = \begin{bmatrix}
		0.7882 & 0.3956 & 0.8333 \\
		0.7062 & 0.7454 & 0.9552 \\
		0.6220 & 0.3116 & 0.4409 \\ 
	\end{bmatrix}\!\!,
	\quad B_i = \begin{bmatrix}
		0.7555 & 0.1557 & 0.3487 \\
		0.1271 & 0.9836 & 0.2030 \\
		0.4735 & 0.4363 & 0.4493 \\ 
	\end{bmatrix}\!\!,
	\quad C_i=0.01\mathds{1}_3^T,\\\label{Case study 1}
	&E_i=\begin{bmatrix}
		0.6482; \!&\!
		0.6008; \!&\!
		0.6209
	\end{bmatrix}\!\!,
	\quad F_i =\begin{bmatrix}
		0.5146; \!&\!
		0.8756; \!&\!
		0.2461
	\end{bmatrix}^T\!\!\!\!\!,
	\quad R_i =\begin{bmatrix}
		0.4974; \!&\!
		0.3339; \!&\!
		0.4527
	\end{bmatrix}\!\!,
\end{align}
for $i\in\{1,2,3,4\}$. The internal input and output matrices are also given by
\begin{align}\notag
	C_{14}=C_{23}=C_{31}=C_{42}= 0.01\mathds{1}_3^T\!,\quad D_{13}=	D_{24}=	D_{32}=D_{41}\!=\!\begin{bmatrix}
		0.074;  \!&\! 
		0.010;  \!&\! 
		0.086  
	\end{bmatrix}\!\!.
\end{align} 
We consider $\varphi_i(x) = sin(x)$, $\forall i\in\{1,\ldots,4\}$. Then functions $\varphi_i$ satisfy condition \eqref{Eq_6a} with $b = 1$.
In the following, we first construct the reduced-order version of the given dynamic by satisfying conditions ~\eqref{Eq11a}-\eqref{Eq16a}. We then establish relations between subsystems by fulfilling condition~\eqref{Eq17a}. Afterwards, we satisfy the compositionality condition~\eqref{compositionality condition} to get a relation on the composed system, and finally, we utilize Theorem~\ref{closeness} to provide the probabilistic closeness guarantee between the interconnected model and its constructed finite MDP.

Conditions~\eqref{Eq11a}-\eqref{Eq16a} are satisfied with, $\forall i\in\{1,2,3,4\}$,
\begin{align}\notag
	Q_i &= \begin{bmatrix}
		-1.6568; \!&\!
		-1.2280;  \!&\! 
		1.9276 
	\end{bmatrix}\!\!,
	\quad S_i= \begin{bmatrix}
		0.0775; \!&\! 
		0.0726;  \!&\! 
		-0.1759 
	\end{bmatrix}\!\!,\\\notag
	P_{i} &= \begin{bmatrix}
		0.5931; \!&\! 
		0.3981;  \!&\! 
		0.5398 
	\end{bmatrix}\!\!, 
	\quad L_{1i} =\begin{bmatrix}
		-0.6546; \!&\! 
		-0.4795;  \!&\! 
		-0.2264  
	\end{bmatrix}\!\!,\\\notag
	L_{2i} &= \begin{bmatrix}
		-0.1713; \!&\!
		-0.0777; \!&\! 
		-0.1044  
	\end{bmatrix}\!\!,\quad P_{wi} = 1,  M_i = \mathds{I}_{3}. 
\end{align}
Accordingly, matrices of reduced-order systems can be obtained as $,\forall i\in\{1,2,3,4\},$
\begin{align}\notag
	&\hat A_{{\textsf r}i} = 0.5127, \hat E_{{\textsf r}i} =0.3,\hat F_{{\textsf r}i} = 0.7866,\hat C_{{\textsf r}i} =0.0371,\hat D_{{\textsf r}i}= 0.1403, \hat R_{{\textsf r}i}= 0.8386.
\end{align}
Moreover, we compute $\tilde R_i=(B_i^TM_iB_i)^{-1} B_i^TM_i P_i\hat B_{{\textsf r}i}$, $i\in\{1,2,3,4\}$, to make chance constraint \eqref{Eq17a} less conservative. By taking $\hat B_{{\textsf r}i} = 2$, we have $\tilde R_i = [
1.1418;   
0.5182;  
0.6965
]$.
The interface functions for $i\in\{1,2,3,4\}$ are acquired by \eqref{Eq17aa} as
\begin{align}\notag
	\nu_i=&\begin{bmatrix}
		-0.6665 \,&\, -0.3652 \,&\, -0.9680 \\
		-0.4372 \,&\, -0.5536 \,&\, -0.5781 \\
		-0.4012 \,&\, -0.1004 \,&\, -0.2612 \\ 
	\end{bmatrix}\!\! 
	(x_i-P_i\hat x_i)+Q_i\hat x_i+\tilde R_i\hat \nu_i + S_i \hat w_i+L_{1i}\varphi_i(F_ix_i)-L_{2i}\varphi_i(F_iP_i \hat x_i).
\end{align}

We proceed with showing that condition~\eqref{Eq17a} holds as well, using Remark~\ref{rem:chi}. This condition can be satisfied via the S-procedure~\cite{boyd2004convex}, which enables us to reformulate \eqref{Eq17a} as existence of $\lambda\ge 0$ such that matrix inequality 
\begin{align}\label{S-procedure}
	\lambda_i \begin{bmatrix}
		\tilde F_{1i} & \tilde g_{1i} \\
		\tilde g_{1i}^T & \tilde h_{1i}  \\
	\end{bmatrix}\!\!
	- \begin{bmatrix}
		\tilde F_{2i} & \tilde g_{2i} \\
		\tilde g_{2i}^T & \tilde h_{2i}  \\
	\end{bmatrix}\! \succeq 0,
\end{align}
holds. Here, $\tilde F_{1i}$ and $\tilde F_{2i}$ are symmetric matrices, $\tilde g_{1i}$ and $\tilde g_{2i}$  are vectors, $\tilde h_{1i}$ and $\tilde h_{2i}$ are real numbers.
We first bound the external input of abstract systems as $\hat \nu_i^2 \leq c_{\hat \nu i}$ and select $c_{\varsigma i} = \mathcal{X}_2^{-1}(1-\delta_i)$, for all $i\in\{1,2,3,4\}$.
Then matrices, vectors and real numbers of inequality~\eqref{S-procedure}, $\forall i\in\{1,2,3,4\}$, can be constructed as in~\eqref{Matrices} and~\eqref{Vectors} provided in the Appendix.
By taking $\epsilon_{i} = 1.25$, $\epsilon_{w_i} = 0.05$, $c_{\hat \nu_ i} = 0.25$, $\delta_i = 0.001$, $\beta_i = 0.1$, $\lambda_i = 0.347$, for all $i\in\{1,2,3,4\}$, 
one can readily verify that the matrix inequality~\eqref{S-procedure} holds. Then $\widehat \Sigma_i$ is ($\epsilon_i, \delta_i $)-stochastically simulated by $\Sigma_i$  with relations
\begin{align}\notag
	&\mathscr{R}_{xi}=\Big\{(x_i,\hat x_i)~|~(x_i - P_i\hat x_i)^T M_i(x_i- P_i
	\hat x_i) \leq \epsilon_i^2\Big\},\quad \mathscr{R}_{wi}=\Big\{(w_i,\hat w_i)~|~(w_i-\hat w_i)^2 \leq \epsilon_{wi}^2\Big\},
\end{align}
for $i\in\{1,2,3,4\}$.
We proceed with showing that the compositionality condition in~\eqref{compositionality condition} holds, as well.  To do so, by employing S-procedure, one should satisfy the  matrix inequality in~\eqref{S-procedure} with the following matrices:

\begin{align}
	&\tilde F_{1i} =\begin{bmatrix}\notag
		M_i & -M_i P_i\\
		* & P_i^T M_i P_i\\
	\end{bmatrix}\!\!,
	\quad\tilde F_{2i} = \begin{bmatrix}\notag
		C_{{\textsf r}i}^T M_{wi} C_{{\textsf r}i} & -C_{{\textsf r}i}^T  M_{wi} P_{wi} \hat C_{{\textsf r}i}\\
		* & \hat C_{{\textsf r}i}^T P_{wi}^T M_{wi} P_{wi}\hat C_{{\textsf r}i}\\
	\end{bmatrix}\!\!,
	\\\notag
	&\tilde g_{1i} = \tilde g_{2i} = \mathbf{0}_4, \quad\tilde h_{1i} = -\epsilon_{i}^2 , \tilde h_{2i} = -\epsilon_{wi}^2,
\end{align}
for $i\in\{1,2,3,4\}$. This condition is satisfiable with $\lambda_i = 0.001 ~ \forall i\in\{1,2,3,4\}$, thus $\widehat \Sigma$ is ($\epsilon, \delta $)-stochastically simulated by $\Sigma$  with $\epsilon = 6$, and $\delta = 0.003$. According to~\eqref{Closness}, we guarantee that the distance between outputs of $\Sigma$ and of $\widehat \Sigma$ will not exceed $\epsilon = 6$ during the time horizon $T_k=10$ with probability at least $96\%$ ($\gamma = 0.04$).

\subsection{Comparison}
To demonstrate the effectiveness of the proposed approach, let us now compare the guarantees provided by our approach and by~\cite{lavaei2018CDCJ, lavaei2017HSCC}. Note that our result is based on the $\delta$-lifted relation while~\cite{lavaei2018CDCJ, lavaei2017HSCC} employ dissipativity-type reasoning
to provide a compositional methodology for constructing both infinite abstractions (reduced-order models) and finite MDPs in two consecutive steps. Since we are not able to satisfy the proposed matrix inequalities in~\cite[Ineqality (22)]{lavaei2017HSCC}, and~\cite[Inequality (5.5)]{lavaei2018CDCJ} for the given system in~\eqref{Case study 1},
we change the system dynamics to have a fair comparison. In other words, in order to show the conservatism nature of the existing techniques in~\cite{lavaei2017HSCC, lavaei2018CDCJ}, we provide another example and compare our techniques with the existing ones in great detail.

The matrices of the new system are given by
\begin{align}\notag
	A_i = \mathds{I}_{5}, ~B_i = \mathds{I}_{5}, ~C_i = 0.05\mathds{1}_{5}^T, ~R_i = \mathds{1}_{5},
\end{align}
for $i\in\{1,2,3,4\}$, where matrices $E_i, F_i$ are identically zero. The internal input and output matrices are also given by:
\begin{align}\notag
	C_{14}=C_{23}=C_{31}=C_{42}=0.05\mathds{1}_{5}^T, \quad D_{13}=D_{24}= D_{32}=D_{41}= 0.1\mathds{1}_{5}.
\end{align}
Conditions~\eqref{Eq11a},\eqref{Eq12a},\eqref{Eq15a},\eqref{Eq16a} are satisfied by:
\begin{align}\notag
	M_i = \mathds{I}_{5}, ~P_{xi} = \mathds{1}_{5},~ P_{wi} = 1, ~Q_i =\mathds{1}_{5}, ~S_i= 0.1\mathds{1}_{5},
\end{align}
for $i\in\{1,2,3,4\}$. Accordingly, the matrices of reduced-order systems are given as:
\begin{align}\notag
	\hat A_{{\textsf r}i} = 2, \hat C_{{\textsf r}i} =0.25, \hat D_{{\textsf r}i}= 0.2, \hat R_{{\textsf r}i}= 0.97, ~\forall i\in\{1,2,3,4\}.
\end{align}
Moreover, by taking $\hat B_{{\textsf r}i} = 1$, we compute $\tilde R_i$, $i\in\{1,2,3,4\}$, as $\tilde R_i = \mathds{1}_{5}$.
The interface function for $i\in\{1,2,3,4\}$ is computed as:
\begin{align}\notag
	\nu_i=-0.95\mathds{I}_{5}(x_i-\mathds{1}_{5}\hat x_i)+\mathds{1}_{5}\hat x_i+\mathds{1}_{5}\hat \nu_i+0.1\mathds{1}_{5}\hat \omega_i.
\end{align}
We proceed with showing that condition~\eqref{Eq17a} holds, as well. By taking 
\begin{align}\notag
	\epsilon_{i} = 5, \epsilon_{w_i} = 0.75,  c_{\hat \nu_i} = 0.25, \delta_i = 0.001, \beta_i = 0.1, \lambda_i = 0.825,\quad  \forall i\in\{1,2,3,4\},
\end{align}
and by employing S-procedure, one can readily verify that condition~\eqref{Eq17a} holds. Then $\widehat \Sigma_i$ is ($\epsilon_i, \delta_i $)-stochastically simulated by $\Sigma_i$,
for $i\in\{1,2,3,4\}$. Additionally, by applying S-procedure, one can readily verify that $\widehat \Sigma$ is ($\epsilon, \delta $)-stochastically simulated by $\Sigma$  with $\epsilon = 20$, and $\delta = 0.005$. According to~\eqref{Closness}, we guarantee that the distance between outputs of $\Sigma$ and of $\widehat \Sigma$ will not exceed $\epsilon = 20$ during the time horizon $T_k=5$ with probability at least $97\%$ ($\gamma = 0.03$). 

Now we apply the proposed results in~\cite{lavaei2017HSCC,lavaei2018CDCJ} for the same matrices of the new system and also employing the same $\epsilon$ and discretization parameter $\beta$. Since the proposed approaches in~\cite{lavaei2017HSCC,lavaei2018CDCJ} are presented in two consecutive steps, we employ the next proposition which provides the overall error bound in two-step abstraction scheme. 

\begin{proposition}\label{proposition}
	Suppose $\Sigma_1$, $\Sigma_2$, and $\Sigma_3$ are three stochastic systems without internal signals. For any external input trajectories $\nu_1$, $\nu_2$, and $\nu_3$ and for any $a_1$, $a_2$, and $a_3$ as the initial states of the three systems, if
	\begin{align}\notag
		&\mathbb{P}\left\{\sup_{0\leq k\leq T_k}\Vert y_{1a_1\nu_1}(k)-y_{2 a_2\nu_2}(k)\Vert\geq\epsilon_1\,|\,[a_1;a_2]\right\}\leq \gamma_1,\\\notag
		&
		\mathbb{P}\left\{\sup_{0\leq k\leq T_k}\Vert y_{2 a_2\nu_2}(k)-y_{3a_3 \nu_3}(k)\Vert\geq\epsilon_2\,|\,[a_2;a_3]\right\}\leq \gamma_2,
	\end{align}
	for some $\epsilon_1,\epsilon_2>0$ and $\gamma_1,\gamma_2\in]0~1[$, then the probabilistic mismatch between output trajectories of $\Sigma_1$ and $\Sigma_3$ is quantified as
	\begin{align}\notag
		\mathbb{P}&\left\{\sup_{0\leq k\leq T_k}\Vert y_{1a_1\nu_1}(k)-y_{3a_3\nu_3}(k)\Vert\geq\epsilon_1+\epsilon_2\,|\,[a_1;a_2;a_3]\right\}\leq \gamma_1+\gamma_2.
	\end{align}	
\end{proposition}
The proof is provided in the Appendix. 

By applying the proposed results in~\cite{lavaei2018CDCJ} to construct the infinite abstraction $\widehat \Sigma_{\textsf r}$, one can guarantee that the distance between outputs of $\Sigma$ and of $\widehat \Sigma_{\textsf r}$ will exceed $\epsilon_1 = 15$ during the time horizon $T_k=5$ with probability at most $87.94\%$, i.e.,
\begin{equation*}
	\mathbb P(\Vert y_{a\nu}(k)-\hat y_{{\textsf r}\hat a_{\textsf r} \hat\nu_{\textsf r}}(k)\Vert\ge 15,\,\, \forall k\in[0,5])\le 87.94\,.
\end{equation*}
After applying the proposed results in~\cite{lavaei2017HSCC} to construct the finite abstraction $\widehat \Sigma$ from $\widehat \Sigma_{\textsf r}$, one can guarantee that the distance between outputs of $\widehat \Sigma_{\textsf r}$ and of $\widehat \Sigma$ will exceed $\epsilon_2 = 5$ during the time horizon $T_k=5$ with probability at most $0.0117\%$, i.e.,
\begin{equation*}
	\mathbb P(\Vert \hat y_{{\textsf r}\hat a_{\textsf r} \hat\nu_{\textsf r}}(k)-\hat y_{\hat a \hat\nu}(k)\Vert\ge 5,\,\, \forall k\in[0,5])\le 0.0117.
\end{equation*}
By employing Proposition~\ref{proposition}, one can guarantee that the distance between outputs of $\Sigma$ and of $\widehat \Sigma$ will exceed $\epsilon = 20$ during the time horizon $T_k=5$ with probability at most $0.8911\%$, i.e.
\begin{equation*}
	\mathbb P(\Vert y_{a\nu}(k)-\hat y_{\hat a \hat\nu}(k)\Vert\ge 20,\,\, \forall k\in[0,5])\le 0.8911.
\end{equation*}
This means that the distance between outputs of $\Sigma$ and of $\widehat \Sigma$ will not exceed $\epsilon = 20$ during the time horizon $T_k=5$ with probability at least $0.1089\%$. As seen, our provided results dramatically outperform the ones proposed in~\cite{lavaei2017HSCC,lavaei2018CDCJ}. More precisely, since our proposed approach here is presented in a unified framework than two-step abstraction scheme which is the case in~\cite{lavaei2017HSCC,lavaei2018CDCJ}, we only need to check our proposed conditions one time, and consequently, our proposed approach here is much less conservative.

\section{Discussion}
In this paper, we provided a unified compositional scheme for constructing both finite and infinite abstractions of gMDPs with internal inputs.
We defined ($\epsilon, \delta$)-approximate probabilistic relations that are suitable for constructing compositional abstractions of gMDPs.
We focused on a specific class of nonlinear dynamical systems, and constructed both infinite (reduced-order models) and finite abstractions in a unified framework, using quadratic relations on the space and linear interface functions.
We then provided conditions for composing such relations.
Finally, we demonstrated the effectiveness of the proposed results by considering a network of four nonlinear systems (totally 12 dimensions) and constructing finite gMDPs from their reduced-order versions (together 4 dimensions) with guaranteed bounds on their probabilistic output trajectories. We benchmarked our results against the compositional abstraction techniques of~\cite{lavaei2017HSCC,lavaei2018CDCJ}, and showed that our proposed approach is much less conservative than the ones proposed in~\cite{lavaei2017HSCC,lavaei2018CDCJ}. 

\section{Acknowledgment}
This work was supported in part by the H2020 ERC Starting Grant AutoCPS (grant agreement No. 804639).

\bibliographystyle{alpha}
\bibliography{biblio}

\section{Appendix}

\begin{definition}\textbf{(\cite{SIAM17})}
	\label{Def: Interconnected}
	Consider two gMDPs without internal inputs $\Sigma =(X,U,\pi, T,Y,h)$ and $\widehat \Sigma =(\hat X,\hat U,\hat \pi, \hat T, Y,\hat h)$, that have the same output spaces. $\widehat \Sigma$ is ($\epsilon, \delta $)-stochastically simulated by $\Sigma$, i.e. $ \widehat \Sigma\preceq_{\epsilon}^{\delta}\Sigma $, if there exists a relation $\mathscr{R}_x\subseteq X \times \hat X$ for which there exists a Borel measurable stochastic kernel $\mathscr{L}_{T}(\cdot~|~ x, \hat x, \hat \nu)$ on $X \times \hat X$ such that
	\begin{itemize}
		\item $\forall (x,\hat x) \in \mathscr{R}_x, ~\Vert h(x)- \hat h (\hat x) \Vert \leq \epsilon$, 
		\item $\forall (x,\hat x) \in \mathscr{R}_x, \forall \hat \nu \in \hat U, \exists \nu \in U~~~$ \!\!\!\!\! such that $T(\cdot~|~ x,\nu(x,\hat x, \hat \nu))~\mathscr{\bar R}_{\delta} 
		~ \hat T(\cdot~|~ \hat x, \hat \nu)$
		with $\mathscr{L}_{T}(\cdot~|~ x, \hat x, \hat \nu)$,
		\item $ \pi ~ \mathscr{\bar R}_{\delta} ~ \hat \pi $.
	\end{itemize}
\end{definition}

{\bf Matrices appeared in~\eqref{S-procedure}:}
\begin{align}\label{Matrices}
	&\tilde F_{1i} = \begin{bmatrix}
		M_i & \mathbf{0}_{3\times 3} &\mathbf{0}_3 & \mathbf{0}_3 & \mathbf{0}_3 & \mathbf{0}_3\\
		\mathbf{0}_{3\times 3}& \mathbf{0}_{3\times 3} &\mathbf{0}_3 & \mathbf{0}_3 & \mathbf{0}_3 & \mathbf{0}_3\\
		* & * & M_{wi} & 0 & 0 & 0 \\
		* & * &* & 1 & 0 & 0\\
		* & * & * &* & 1 & 0\\
		* & * & * & * &* & 1\\
	\end{bmatrix}\!\!,
	\quad \tilde F_{2i} = \begin{bmatrix}
		\tilde F_{11i} & \tilde F_{12i} & \tilde F_{13i} & \tilde F_{14i} & \tilde F_{15i}& \tilde F_{16i}\\
		* & \tilde F_{22i} & \tilde F_{23i} & \tilde F_{24i} & \tilde F_{25i}& \tilde F_{26i} \\
		* & * & \tilde F_{33i} & \tilde F_{34i} & \tilde F_{35i}& \tilde F_{36i}\\
		* & * & * & \tilde F_{44i} & \tilde F_{45i}& \tilde F_{46i}\\
		* & * & * & * & \tilde F_{55i}& \tilde F_{56i}\\
		* & * & * & * & *& \tilde F_{66i}\\
	\end{bmatrix}\!\!,
\end{align}
where
\begin{align}\notag
	&\tilde F_{11i} \!=\! (A_i\!+\!B_iK_i)^T M_i(A_i\!+\!B_iK_i), \tilde F_{12i} \!=\!(A_i\!+\!B_iK_i)^T M_i(B_iL_{1i}\!+\!E_i)F_i,\tilde F_{13i} \!=\!(A_i\!+\!B_iK_i)^T M_iD_i,\\\notag
	&\tilde F_{14i} \!=\!(A_i\!+\!B_iK_i)^TM_i(B_i \tilde R_i\!-\!P_i \hat B_{{\textsf r}i}), \tilde F_{15i} \!=\!  (A_i\!+\!B_iK_i)^T M_i P_i,\tilde F_{16i} \!=\! (A_i\!+\!B_iK_i)^T M_i (R_i\!-\!P_i \hat R_{{\textsf r}i}),\\\notag
	&\tilde F_{22i}\!=\! F_i^T(B_iL_{1i}\!+\!E_i)^TM(B_iL_{1i}\!+\!E_i)F_i,\tilde F_{23i}\!=\! F_i^T(B_iL_{1i}\!+\!E_i)^TM_iD_i,\tilde F_{24i}\!=\!  F_i^T(B_iL_{1i}\!+\!E_i)^TM_i\\\notag
	&(B_i \tilde R_i\!-\!P_i \hat B_{{\textsf r}i}),\tilde F_{25i}\!=\! F_i^T(B_iL_{1i}\!+\!E_i)^T M_i P_i, \tilde F_{26i}\!=\! F_i^T(B_iL_{1i}+E_i)^TM_i(R_i\!-\!P_i \hat R_{{\textsf r}i}),\tilde F_{33i}\!=\! D_i^T  M_iD_i,\\\notag
	&\tilde F_{34i}\!=\! D_i^T\! M_i(B_i \tilde R_i\!-P_i \hat B_{{\textsf r}i}),\tilde F_{35i}\!=\! D_i^T  M_i P_i, \tilde F_{36i}\!=\! D_i^T  M_i(R_i\!-\!P_i \hat R_{{\textsf r}i}),\tilde F_{44i}=(B_i \tilde R_i\!-\!P_i \hat B_{{\textsf r}i})^T M_i\\\notag
	&(B_i \tilde R_i -P_i \hat B_{{\textsf r}i}), \tilde F_{45i}\!=\!(B_i \tilde R_i\!-\!P_i \hat B_{{\textsf r}i})^TM_i P_i,\tilde F_{46i}\!=\! (B_i \tilde R_i\!-\!P_i \hat B_{{\textsf r}i})^T M_i (R_i \!-\! P_i \hat R_{{\textsf r}i}),\tilde F_{55i}\!=\! P_i^T M_i P_i,\\\notag
	&\tilde F_{56i}\!=\! P_i^T M_i  (R_i \!-\! P_i \hat R_{{\textsf r}i}),\tilde F_{66i}\!=\! (R_i \!-\! P_i \hat R_{{\textsf r}i})^TM_i (R_i \!-\! P_i \hat R_{{\textsf r}i}).
\end{align}
{\bf Vectors and real numbers appeared in~\eqref{S-procedure}:}

\begin{align}
	\tilde g_{1i} = \tilde g_{2i} = \mathbf{0}_{10}, \quad \tilde h_{1i} = -(\epsilon_{i}^2 + \epsilon_{wi}^2 + c_{\hat \nu i} + c_{\varsigma i} + \beta_i), \tilde h_{2i} = -\epsilon_{i}^2.
\end{align}\label{Vectors}

\begin{proof}\textbf{(Theorem~\ref{closeness})}
	The definition of lifting implies that the initial states of the two systems are in the relation with probability at least $1-\delta$. Moreover, if the two states are in the relation at time $k$, they remain in the relation at time $k+1$ with probability at least $1-\delta$. Then, we can write
	\begin{equation*}
		\mathbb{P}\{(x(k), \hat x(k)) \in \mathscr{R}_{x} ~\text{for all} ~k\in[0,T_k]\}\geq (1-\delta)^{T_k+1}.
	\end{equation*} 
	This can be proved by induction and conditioning the probability on the intermediate states.
	
	Note that if $ \{\hat h (\hat x(k)) \}_{0:T_k} \in \mathsf A^{-\epsilon}$ and $(x(k), \hat x(k)) \in \mathscr{R}_{x}$ for all $k\in[0,T_k]$, then $\{y(k)\}_{0:T_k}\in \mathsf{A}$. As a consequence
	\begin{align}\notag
		\mathbb{P}&\{\{\hat h(\hat x)\}_{0:T_k}\in \mathsf{A}^{-\epsilon}\} \wedge (x(k), \hat x(k)) \in \mathscr{R}_{x} ~\text{for all} ~k\in[0,T_k]\}\leq \mathbb{P}\{\{h(x)\}_{0:T_k}\in \mathsf{A}\}.
	\end{align}
	Now by employing the union bounding argument, we have
	\begin{align}\notag
		& \mathbb{P}\{\{\hat h(\hat x)\}_{0:T_k}\in \mathsf{A}^{-\epsilon}\}-(1-\delta)^{T_k+1}\leq \mathbb{P}\{\{\hat h(\hat x)\}_{0:T_k}\in \mathsf{A}^{-\epsilon}\wedge (x(k), \hat x(k)) \in \mathscr{R}_{x},\text{for all} ~k\in[0,T_k]\}.
	\end{align}
	Then
	\begin{align}\notag
		1&-\mathbb{P}\{\{\hat h(\hat x)\}_{0:T_k}\in \mathsf{A}^{-\epsilon}\wedge (x(k), \hat x(k)) \in \mathscr{R}_{x} ~\text{for all} ~k\in[0,T_k]\}\\\notag
		&\leq(1-\mathbb{P}\{\{\hat h(\hat x)\}_{0:T_k}\in \mathsf{A}^{-\epsilon}\})+(1-\mathbb{P}\{(x(k), \hat x(k)) \in \mathscr{R}_{x}~\text{for all} ~k\in[0,T_k]\})\\\notag
		&\leq(1-\mathbb{P}\{\{\hat h(\hat x)\}_{0:T_k}\in \mathsf{A}^{-\epsilon}\})+(1-(1-\delta)^{T_k+1}).
	\end{align}
	One can deduce that 
	\begin{align}\notag
		\mathbb{P}\{\{\hat h(\hat x)\}_{0:T_k}\in \mathsf{A}^{-\epsilon}\}-(1-(1-\delta)^{T_k+1})\leq\mathbb{P}\{\{h(x)\}_{0:T_k}\in \mathsf{A}\}.
	\end{align}
	Similarly, if $\{h(x(k))\}_{0:T_k}\in \mathsf{A}$ and $(x(k),\hat x(k)) \in \mathscr{R}_{x}$, then $\{\hat h(\hat x(k))\}_{0:T_k}\in \mathsf{A}^\epsilon$. Thus via similar arguments it holds that
	\begin{align}\notag
		\mathbb{P}\{\{h(x)\}_{0:T_k}\in \mathsf{A}\}\leq\mathbb{P}\{\{\hat h(\hat x)\}_{0:T_k}\in \mathsf{A}^{\epsilon}\} +(1-(1-\delta)^{T_k+1}).
	\end{align}
	
\end{proof}	

\begin{proof}\textbf{(Theorem~\ref{Thm1})}
	We first show that the first condition in Definition~\ref{Def: Interconnected} holds. For any $x=[{x_1;\ldots;x_N}]\in X$ and  $\hat x=[{\hat x_1;\ldots;\hat x_N}]\in \hat X$ with $x \mathscr{R}_x \hat x$, one gets:
	\begin{align}\notag
		\Vert h(x)-\hat h(\hat x) \Vert&=\Vert [h_{11}(x_1);\ldots;h_{NN}(x_N)]-[\hat h_{11}(\hat x_1);\ldots;\hat h_{NN}(\hat x_N)]\Vert\\\notag
		&
		\le\sum_{i=1}^N \Vert  h_{ii}(x_i)-\hat h_{ii}(\hat x_i) \Vert\le \sum_{i=1}^N \Vert  h_{i}(x_i)-\hat h_{i}(\hat x_i)\Vert\le \sum_{i=1}^N \epsilon_{i}.
	\end{align} 
	As seen, the first condition in Definition~\ref{Def: Interconnected} holds with $\epsilon = \sum_{i=1}^N \epsilon_{i}$.
	The second condition is also satisfied as follows.
	For any $(x,\hat x)\in \mathscr{R}_x$, and $\hat\nu\in\hat U$, we have:
	\begin{align*}
		& \mathscr L\Big\{x'\mathscr{R}_x \hat x'\,|\, x,\hat x,\hat\nu\Big\} = \mathscr L\Big\{x_i'\mathscr{R}_{x_i} \hat x_i',\,\, i \in \{1,2,\ldots, N\}\,|\, x,\hat x,\hat\nu\Big\} \\
		& = \prod_{i=1}^{N}\mathscr L_i\Big\{x_i'\mathscr{R}_{x_i} \hat x_i' ,|\, g_i(x),\hat g_i(x),\hat\nu_i\Big\} \geq \prod_{i=1}^{N}(1-\delta_i).
	\end{align*} 
	The second condition in Definition~\ref{Def: Interconnected} also holds with $\delta = 1-\prod_{i=1}^{N}(1-\delta_i)$ which completes the proof.
	%	\qed
\end{proof}

\begin{proof}\textbf{(Theorem~\ref{Thm2})}
	First, we show that the first condition in Definition~\ref{Def: subsystems}  holds for all $(x,\hat x) \in \mathscr{R}_x$.	According to \eqref{Eq11a} and \eqref{Eq12a}, we have
	\begin{equation*}
		\Vert Cx-\hat C_{\textsf r}\hat x\Vert^2=(x-P\hat
		x)^TC^TC(x-P\hat x)\le (x-P\hat
		x)^T M (x-P\hat x)\le
		\epsilon^2\!,
	\end{equation*} 
	for any $(x,\hat x) \in \mathscr{R}_x$. Now we proceed with showing the second condition. This condition requires that $\forall (x,\hat x) \in \mathscr{R}_x, \forall (w,\hat w) \in \mathscr{R}_w, \forall \hat \nu \in \hat U$, the next states $(x',\hat x')$ should also be in relation $\mathscr{R}_x$ with probability at least $ 1-\delta$:
	$$\mathbb{P}\{(x'-P\hat x')^TM(x'-P\hat x')\leq \epsilon^2\} \geq 1-\delta.$$
	Given any $x$, $\hat x$, and $\hat \nu$, we choose $\nu$ via the following \emph{interface} function:
	\begin{align}\label{Eq17aa}
		\nu=&\nu_{\hat \nu}(x,\hat x, \hat w, \hat \nu):=K(x-P\hat x)+Q\hat x+\tilde R\hat \nu+S\hat w+L_1\varphi(Fx)-L_2\varphi(FP \hat x).
	\end{align}
	By substituting dynamics of $\Sigma$ and $\widehat\Sigma$, employing \eqref{Eq13a}-\eqref{Eq16a}, and the definition of the interface function \eqref{Eq17aa}, we simplify
	\begin{align*}
		x'- P\hat x' &= Ax+E\varphi(Fx)+Dw +B\nu_{\hat \nu}(x,\hat x, \hat w, \hat \nu)+R\varsigma \\\notag
		&-P(\hat A_{\textsf r} \hat x+\hat E_{\textsf r}\varphi(\hat F_{\textsf r}x)+\hat D_{\textsf r} \hat w+\hat B_{\textsf r} \hat \nu+\hat R_{\textsf r} \varsigma)+PG,
	\end{align*}
	to
	\begin{align}\notag
		(A&+BK)(x-P\hat x)+D(w-P_w\hat w)+(B\tilde R-P\hat B_{\textsf r})\hat \nu\\\label{Eq_118a}
		&+(BL_1+E)(\varphi(Fx)-\varphi(FP\hat x_{\textsf r}))+(R-P\hat R_{\textsf r})\varsigma + PG,
	\end{align}
	with $G = \hat A_{\textsf r} \hat x+\hat E_{\textsf r}\varphi(\hat F_{\textsf r}\hat x)+\hat D_{\textsf r} \hat w+\hat B_{\textsf r} \hat \nu+\hat R_{\textsf r} \varsigma- \Pi_x(\hat A_{\textsf r} \hat x+\hat E_{\textsf r}\varphi(\hat F_{\textsf r}\hat x)+\hat D_{\textsf r} \hat w+\hat B_{\textsf r} \hat \nu+\hat R_{\textsf r} \varsigma)$. 
	From the slope restriction~\eqref{Eq_6a}, one obtains
	\begin{align}\label{Eq_119a}
		\varphi(Fx)-\varphi(FP\hat x)=\bar\delta(Fx-FP\hat x)=\bar\delta F(x-P\hat x),
	\end{align}
	where $\bar\delta$ is a function of $x$ and $\hat x$, and takes values in the interval $[0,b]$. Using~\eqref{Eq_119a}, the expression in~\eqref{Eq_118a} reduces to
	\begin{align}\notag
		((A+BK)&+\bar\delta(BL_1+E)F)(x-P\hat x)+D(w-P_w\hat w)+(B\tilde R-P\hat B_{\textsf r})\hat \nu+(R-P\hat R_{\textsf r})\varsigma +PG.
	\end{align}
	This gives condition \eqref{Eq17a} for having the probabilistic relation.
\end{proof}

\begin{proof}\textbf{(Proposition~\ref{proposition})}
	By defining
	\begin{align}\notag
		\mathcal{A} &= \{\Vert y_{1a_1\nu_1}(k)-y_{2a_2\nu_2}(k)\Vert<\epsilon_1 \,|\,[a_1;a_2;a_3]\},\\\notag
		\mathcal{B} &= \{\Vert y_{2a_2\nu_2}(k)-y_{3a_3\nu_3}(k)\Vert<\epsilon_2 \,|\,[a_1;a_2;a_3]\},\\\notag
		\mathcal{C} &= \{\Vert y_{1a_1\nu_1}(k)-y_{3a_3\nu_3}(k)\Vert<\epsilon_1+\epsilon_1 \,|\,[a_1;a_2;a_3]\},
	\end{align}
	we have $\mathbb{P}\{\mathcal{\bar A}\}\leq \gamma_1$ and  $\mathbb{P}\{\mathcal{\bar B}\}\leq \gamma_2,$ where $\mathcal{\bar A}$ and $\mathcal{\bar B}$ are the complement of $\mathcal{A}$ and $\mathcal{B}$, respectively. Since $\mathbb{P}\{\mathcal{A}\cap \mathcal{B}\}\leq \mathbb{P}\{\mathcal{C}\}$, we have
	\begin{align}\notag
		\mathbb{P}\{\mathcal{\bar C}\} \leq \mathbb{P}\{\mathcal{\bar A}\cup \mathcal{\bar B}\} \leq \mathbb{P}\{\mathcal{\bar A}\} + \mathbb{P}\{\mathcal{\bar B}\} \leq \gamma_1 + \gamma_2.
	\end{align}
	Then
	\begin{align}\notag
		\mathbb{P}&\left\{\sup_{0\leq k\leq T_k}\Vert y_{1a_1\nu_1}(k)-y_{3a_3\nu_3}(k)\Vert\geq\epsilon_1+\epsilon_2\,|\,[a_1;a_2;a_3]\right\}\leq \gamma_1+\gamma_2.
	\end{align}	
\end{proof}

\end{document}